\documentclass[aps,10pt,prl,twocolumn]{revtex4-1}
\usepackage[utf8]{inputenc}
\usepackage[final]{graphicx}
\usepackage{bm}
\usepackage{times}
\usepackage{verbatim}
\usepackage{xcolor}
\usepackage{amsmath}
\usepackage{mathtools}
\usepackage{amssymb}
\usepackage{paralist}
\usepackage[colorlinks=false,pdfborder={0 0 0}]{hyperref}
\usepackage{t1enc}
\usepackage{graphicx}

\newcommand{\beq}{\begin{equation}}
\newcommand{\eeq}{\end{equation}}
\newcommand{\beqa}{\begin{eqnarray}}
\newcommand{\eeqa}{\end{eqnarray}}
\newcommand{\mbr}{\mathbf{r}}
\newcommand{\mbp}{\mathbf{p}}

\newcommand{\mc}{\mathcal}
\newcommand{\pt}{p_{\theta}}
\newcommand{\pp}{p_{\phi}}
\newcommand{\dpm}{\,\mathrm{d}p}
\newcommand{\dpr}{\,\mathrm{d}p_r}
\newcommand{\dpt}{\,\mathrm{d}p_{\theta}}
\newcommand{\dpp}{\,\mathrm{d}p_{\phi}}
\newcommand{\dr}{\,\mathrm{d}r}
\newcommand{\dth}{\,\mathrm{d}\theta}
\newcommand{\dph}{\,\mathrm{d}\phi}
\newcommand{\dll}{\,\mathrm{d}(\ell^2)}
\newcommand{\deps}{\,\mathrm{d}\mathcal{E}}

\newcommand{\ii}{\'{\i}}

\begin{document}

\title{Nonequilibrium stationary states of 3D self-gravitating systems}
\indent

\author{Fernanda P. C. Benetti}
\author{Ana C. Ribeiro-Teixeira}
\author{Renato Pakter}
\author{Yan Levin}
\affiliation{Instituto de F\ii sica, Universidade Federal do Rio Grande do Sul\\
Caixa Postal 15051, CEP 91501-970, Porto Alegre, RS, Brazil
}

\begin{abstract}

Three dimensional self-gravitating systems do not evolve to thermodynamic equilibrium,
but become trapped in nonequilibrium quasistationary states. In this Letter we 
present a theory which allows us to  
{\it a priori} predict the particle distribution in a
final quasistationary state 
to which a self-gravitating system will evolve from an initial condition which 
is isotropic in particle velocities and satisfies
a virial constraint $2 K=-U$, where $K$ is the total
kinetic energy and $U$ is the potential energy of the system.

\end{abstract}


\maketitle

Unlike systems with short-range forces which relax to thermodynamic equilibrium starting from an arbitrary
initial condition, systems 
with long-range interactions become trapped in nonequilibrium quasistationary states (QSS) the lifetime of 
which diverges with the number of particles \cite{Cha1941,KonKan1992,LatRap1998,LatRap2000,YamBar2004,JaiBou2007,AntFan2007,GabJoy2010a,Cha2013}. For 
interaction potentials unbounded from above, 
the QSS have been observed to have a characteristic core-halo structure \cite{LevPak2014}. 
The extent of the halo is determined by the parametric resonances 
which arise from the collective density oscillations
during the relaxation process~\cite{Glu1994}. The dynamics of 3D self-gravitating systems, however, is significantly more complex due to the existence of unbound states \cite{Pad1990,LevPak2008}. 
Indeed, Newton's gravitational potential is bounded from above, so that the parametric resonances
may actually transfer enough energy to allow some particles to completely escape 
from the gravitational cluster~\cite{LevPak2008,JoyMar2009}.
This makes the study of 3D self-gravitating systems particularly challenging~\cite{ChaBou2005,ChaAlas14}.  
Recently, however, it was shown that if the initial particle distribution function is isotropic in velocity 
and satisfies the, so called,
virial condition (VC), density oscillations and parametric resonances will be suppressed \cite{TelLev2010,TelLev2011,JoyWor2011,BenTel2012}.  The relaxation to equilibrium will then proceed adiabatically. In the thermodynamic limit, each particle of the gravitational cluster will evolve under the action of a quasistatic mean-field potential and the phase-mixing of particle trajectories will lead to a nonequilibrium QSS. In this Letter we will show that it is possible to {\it a priori} predict the density and the velocity distribution functions within the QSS to which a 3D gravitational system will evolve if the
initial distribution is isotropic in particle velocities and satisfies VC.

The virial theorem requires that a stationary gravitational system
must have $2 K=-U$, where $K$ is the total
kinetic energy and $U$ is the potential energy.  This, however 
does not mean that an arbitrary initial distribution which satisfies
the VC will remain stationary.  To be stationary, a distribution
function must be a time-independent solution of the collisionless Boltzmann
(Vlasov) equation \cite{BraHep1977,CamDau2009,RocAma2014}. 
From Jeans' theorem, this will only be the case if the distribution
depends on the phase space coordinates solely through the integrals of motion \cite{BinTre2008}.
Recently, however, it was shown that if the initial particle
distribution $f_0(\mbr,\mbp)$ is spherically symmetric and isotropic in
velocity, $f_0(\mbr,\mbp)=f_0(r,p)$, and satisfies the VC, strong
density oscillations will be suppressed and the relaxation to QSS
will be intrinsically different than for initial distributions which do
not satisfy VC \cite{Yam2008,LevPak2014}. In principle,
a spherically symmetric distribution does not need to be a function of the modulus
of momentum.  A spherical symmetry is compatible with the distribution being a function of both radial and angular momentum independently. 
The assumption of isotropicity is 
included to prevent the radial orbit instability (ROI) which leads to spontaneous
symmetry breaking of the  distribution function.  ROI can occur when
kinetic energy of the system is dominated by the radial velocity component \cite{AguMer1990, BarLan2009}. On the other hand, for isotropic
velocity distributions symmetry breaking 
occur only when the initial distribution deviates strongly from VC \cite{PakMar2013}.
For initial particle distributions isotropic in velocity and satisfying the VC, relaxation to equilibrium is a consequence of phase mixing of particle trajectories \cite{RibBen2014}, while for
non-virial initial conditions
relaxation results from 
excitation of parametric resonances~\cite{Glu1994} and a nonlinear Landau damping \cite{Vil2014,LevPak2014}. 
 
Consider a spherically symmetric --- in both positions and velocities --- initial phase space particle distribution. We will work in the thermodynamic limit $N \rightarrow \infty$, $m \rightarrow 0$, while $mN=M$, where $N$ is the total number of particles, $m$ is the mass of each particle, and $M$ is the total mass of the gravitational system. At $t=0$ the particles are distributed in accordance with the initial distribution $f_0(r,p)$ inside an infinite 3D configuration space. We would like to predict the distribution function
for the system when it relaxes to a QSS. It is easy to see that in the thermodynamic limit the positional correlations  between the particles vanish and all the dynamics is controlled by the mean-field potential \cite{RocAma2014}. Furthermore, if the initial distribution is such that the VC is satisfied, the mean-field potential should vary adiabatically
and the energy of each particle should change little.  
Since the mean-field potential 
is a nonlinear function of position, the particles on the energy shell $[\mc{E},\mc{E}+d\mc{E}]$ with slightly distinct one-particle energies $\mc{E}$ will have incommensurate orbital frequencies. This means that after a transient period, the phase-mixing will result in a uniform particle distribution over the energy shell. The particle distribution in the final QSS can then be obtained by a coarse-graining of the initial distribution 
over the phase space available to the particle dynamics, taking into account
the conservation of the angular momentum of each particle, given the spherical symmetry of the mean-field potential.  

Consider an arbitrary initial particle  distribution 
$f_0(r,p)$ that satisfies VC. For $t>0$ the
particles will evolve under the action of an external adiabatically varying potential $\varphi(r,t)$ which will eventually converge to  some $\psi(r)$. 
Our approach will be to construct a coarse-grained distribution for particles evolving directly
under the action of the {\it static} potential  $\psi(r)$ which will then be calculated self-consistently 
\cite{LeoVan2009,BuyMuk2011,BuyMuk2011a}.  
Clearly such an approximation will only work if the 
variation of $\varphi(r,t)$ is adiabatic and no resonances are excited. This is precisely the case for the initial distributions which are isotropic in velocity and 
satisfy VC \cite{RibBen2014}. 

Since $\psi(r)$ is static and spherically symmetric, 
the energy and the angular momentum of each particle will be preserved. The 
nonlinearity of $\psi(r)$ will lead to phase-mixing of particle trajectories 
with the same energy and angular momentum. 
The number of particles with energy between $[\mc{E},\mc{E}+d\mc{E}]$ 
and the square of the angular momentum between $[\ell^2, \ell^2+d \ell^2]$ is $n(\mc{E},\ell^2) d\mc{E} d \ell^2$ and is conserved throughout dynamics. 
In the QSS these particles will spread over the phase space volume $g(\mc{E},\ell^2) d\mc{E} d \ell^2$, so that 
the coarse-grained distribution function for the QSS will be 
%%%%%%%
\beq\label{eq:fimss_ng}
f(\mc{E},\ell^2)=\frac{n\!\left(\mc{E},\ell^2\right)}{g\!\left(\mc{E},\ell^2\right)}
\eeq
%%%%%%%
The self-consistent  potential $\psi(r)$ must satisfy the Poisson equation,
\beq
\frac{1}{r^2}\frac{\mathrm{d}}{\mathrm{d}r}\left(r^2\frac{\mathrm{d}\psi}{\mathrm{d}r}\right)=4\pi Gm \rho(r)
\eeq
where
\beq 
\rho({\bf r})=\int d^3 p \,\,\,f\left[\mc{E}({\bf r},{\bf p}),\ell^2({\bf r},{\bf p})\right]
\eeq
is the asymptotic particle density. This gives
us a closed set of equations which can be used to calculate the distribution
function in the QSS.  To simplify the notation we will scale all the 
distances to an arbitrary length scale $L_0$, time to $\sqrt{L_0^3/GM}$, the potential to $GM/L_0$,
and the energy to $GM^2/L_0$.

Because of the conservation of the angular momentum of each particle, 
it is convenient to work with the canonical positions $(r,\theta,\phi)$ and  conjugate momenta $(p_r,\pt,\pp)$.  
Note that in terms of these variables 
the invariant phase space measure is   $\mathrm{d}^3x\mathrm{d}^3p=\dr\dth\dph\,\dpr\dpt\dpp$. The particle energy and square modulus of the 
angular momentum are
%%%%%%%
\begin{align}
\epsilon(r,\theta,p_r,\pt,\pp)&=\frac{1}{2}\left(p_r^2+\frac{\pt^2}{r^2}+\frac{\pp^2}{r^2\sin^2\theta}\right)+\psi(r),\label{eq:epsilon}\\
l^2(\theta,\pt,\pp)&=\pt^2+\frac{\pp^2}{\sin^2\theta}\label{eq:am}
\end{align}
%%%%%%%
respectively. 
The density of states $g(\mc{E},\ell^2)$ is
%%%%%%%
\begin{align}\label{eq:ggen}
g(\mc{E},\ell^2)&=\int\!\! \dpr \dpt \dpp\! \int\!\! \dr \dth \dph
\,\,\delta\left[\ell^2\!-l^2(\theta,\pt,\pp)\right]\nonumber\\
&\quad\times\delta\left[\mc{E}-\epsilon(r,\theta,p_r,\pt,\pp)\right]
\end{align}
%%%%%%%
and the particle phase space density $n(\mc{E},\ell^2)$ is
%%%%%%%
\begin{align}\label{eq:ngen}
n(\mc{E}&,\ell^2)=\int \dpr \dpt \dpp \int \dr \dth \dph
\,\delta\!\left[\ell^2-l^2(\theta,\pt,\pp)\right]\nonumber\\
&\times\delta\!\left[\mc{E}-\epsilon(r,\theta,p_r,\pt,\pp)\right]\nonumber\\
&\times f_0\!\left(r,\sqrt{p_r^2+\frac{\pt^2}{r^2}+\frac{\pp^2}{r^2\sin^2\theta}}\right).
\end{align}
%%%%%%%
Integration over all the variables in Eqs. \eqref{eq:ggen} and \eqref{eq:ngen}, other than $\dr$, can be performed with the help of a Dirac delta function identity 
\beq \label{Dirac}
\delta[f(x)]=\frac{\sum_i \delta(x-x_i)}{\vert f'(x_i) \vert}, 
\eeq
where $x_i$ is the $i$'th root of $f(x)$. Carrying out the integration we obtain
the coarse-grained distribution function for the QSS,
%%%%%%%
\beq\label{eq:fel}
f(\mc{E},\ell^2)=\frac{\int \dr \,f_0\!\left[r,\sqrt{2\big(\mc{E}-\psi(r)}\big)\right]\frac{\Theta\left[\mc{E}-\frac{\ell^2}{2r^2}-\psi(r)\right]}{\sqrt{\mc{E}-\frac{\ell^2}{2r^2}-\psi(r)}}}{\int \dr\,\frac{ \Theta\left[\mc{E}-\frac{\ell^2}{2r^2}-\psi(r)\right]}{\sqrt{\mc{E}-\frac{\ell^2}{2r^2}-\psi(r)}}},
\eeq
%%%%%%%
where $\Theta$ is the Heaviside step function.  
The coarse-grained distribution function depends on position and momentum 
only through the conserved quantities $\mc{E}$ 
and $\ell^2$; therefore, it is automatically a stationary solution of the Vlasov equation.
%, since $\{f(\mc{E},\ell^2),\mathcal{H}\}=0$}.  

The Poisson equation can be rewritten as 
%%%%%%%
\beq\label{eq:poisson}
r^2\frac{\mathrm{d}^2\psi}{\mathrm{d}r^2}+2r\frac{\mathrm{d}\psi}{\mathrm{d}r}=N(r)
\eeq
%%%%%%%
where $N(r)=4\pi r^2\rho(r)$, or
%%%%%%%
\beq\label{eq:nr}
N(r)=\int\!\!\dpr\dpt\dpp\!\int\!\!\dth\dph\, f\!\left(\mc{E},\ell^2\right).
\eeq
%%%%%%%
%It is convenient to rewrite Eq \eqref{eq:nr} in terms of $f(\mc{E},\ell)$, which we do know. Then,
%%%%%%%
%\beq\label{eq:nr1}
%N(r)=\int\dpr\dpt\dpp\int\dth\dph\, f(\mc{E},\ell)=\int\dpr\dpt\dpp\int\dth\dph f\left[\frac{1}{2m}\left(p_r^2+\frac{\pt^2}{r^2}+\frac{\pp^2}{r^2\sin^2\theta}\right)+\psi(r),\pt^2+\frac{\pp^2}{\sin^2\theta}\right].
%\eeq
%%%%%%%
Multiplying Eq \eqref{eq:nr} by the identity
%%%%%%%%%%%%%%%% 
\beq \label{eq:id1}
\int \dll\, \delta\!\left(\ell^2-\pt^2-\frac{\pp^2}{\sin^2\theta}\right)=1 \,,
\eeq 
%%%%%%%%%%%%%%%%
and changing the order of integration, we can write
%%%%%%%
\begin{align} \label{eq:nr2}
N(r)&=\int\!\!\dll\dpr\dpt\dpp\!\int\!\!\dth\dph
\,\,\delta\!\left(\ell^2-\pt^2-\frac{\pp^2}{\sin^2\theta}\right)\nonumber\\
&\times f\!\left(\frac{p_r^2}{2}+\frac{\pt^2}{2r^2}+\frac{\pp^2}{2r^2\sin^2\theta}
+\psi(r),\pt^2+\frac{\pp^2}{\sin^2\theta}\right) \,.
\end{align}
%%%%%%%
The integration over the variables $\pt$, $\pp$, $\theta$, $\phi$ can now be 
performed explicitly with the help of Eq. \eqref{Dirac}. 
Finally, changing the integration variable from $p_r$ to $\mc{E}$,
%%%%%%%
%\begin{align*}
%p_r^2&=2m\left(\mc{E}-\frac{\ell^2}{2mr^2}-\psi(r)\right),\\
%\dpr&=\frac{m}{p_r}\deps,
%\end{align*}
%%%%%%%
Eq. \eqref{eq:nr2} simplifies to
%%%%%%%
\beq \label{eq:nr3}
N(r)=8\pi^2\int\limits_0^{\infty}\dll\int\limits_{\mc{E}_0}^{\infty}\deps\,f\left(\mc{E},\ell^2\right) \frac{\Theta\left[\mc{E}-\frac{\ell^2}{2r^2}-\psi(r)\right]}{\sqrt{2\left(\mc{E}-\frac{\ell^2}{2r^2}-\psi(r)\right)}}
\eeq
%%%%%%%
where the lower limit of integration is $\mc{E}_0=\frac{\ell^2}{2r^2}+\psi(r)$ and 
 $f\left(\mc{E},\ell^2\right)$ is given by Eq \eqref{eq:fel}.  Substituting Eq. \eqref{eq:nr3} into 
Eq. \eqref{eq:poisson}, we find an integro-differential equation for the
gravitational potential $\psi(r)$ in the QSS.  Eq. \eqref{eq:poisson} can be solved numerically using Picard iteration.  Once the gravitational potential is
known, the coarse-grained distribution function  can be easily 
calculated by performing the integration in Eq. \eqref{eq:fel}. 

We next validated the proposed theory by comparing the marginal
position and velocity distribution functions $N(r)$ and $N(p)$ to explicit molecular dynamics (MD) simulations of a 
3D self-gravitating system of $N$ particles. The simulations were performed using 
a version of particle-in-cell (PIC) algorithm, in which each particle interacts with
a mean-field potential produced by all other particles \cite{LevPak2014}.  
In the absence of ROI this simulations produce identical particle distributions in QSS as 
calculated using traditional binary interaction methods, but are three orders of
magnitude faster.  This allows us to easily reach the QSS~\cite{SupMatPRL}. 
The density distribution $N(r)$ is
given by Eq \eqref{eq:nr3}. To obtain the momentum distribution we first calculate the
distribution 
%%%%%%%
\beq\label{eq:npr}
N(p_r)=\int\!\!\dr\dpt\dpp\!\int\!\!\dth\dph\, f\!\left(\mc{E},\ell^2\right) \,,
\eeq
%%%%%%%
where $\mc{E}=p_r^2/2+\frac{\ell^2}{2r^2}+\psi(r)$ and $\ell^2=\pt^2+\frac{\pp^2}{\sin^2\theta}$.
The change of variable from $p_r$ to the modulus of momentum $p$ can be performed with the
help of Eq. \eqref{eq:id1} and the  identity 
\beq
\int \dpm^2 \, \delta\!\left(p^2- p_r^2 -\frac{\ell^2}{r^2} \right)=1
\eeq 
yielding,
\beq \label{eq:np}
N(p)=8\pi^2 p\int\limits_0^{\infty}\dll\int\limits_{0}^{\infty}\dr\,f\left(\mc{E},\ell^2\right) \frac{\Theta\left[p^2-\frac{\ell^2}{r^2}\right]}{\sqrt{p^2-\frac{\ell^2}{r^2}}} \,
\eeq
%%%%%%%
where $\mc{E}=p^2/2 +\psi(r)$.
%This is a QSS analogy of the Maxwell-Boltzmann velocity distribution for systems
%with short range interactions.

\begin{figure}
\begin{center}
\includegraphics[width=0.45\textwidth]{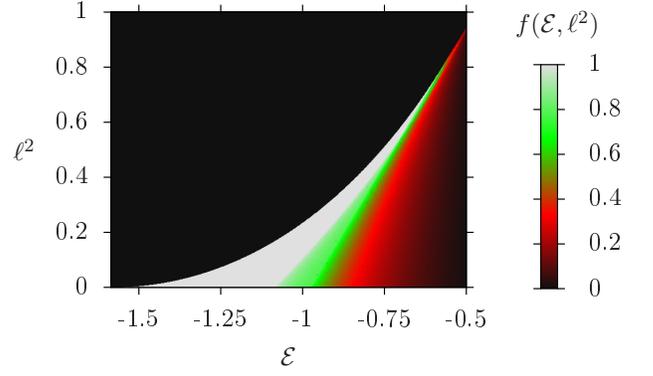}
\caption{Distribution function in energy and angular momentum for the QSS, 
for an initial waterbag distribution, Eq \eqref{eq:wbdist}, satisfying VC. \label{fig:fel_wb}}
\end{center}
\end{figure}

We first consider a waterbag initial distribution, 
%%%%%%%
\beq\label{eq:wbdist}
f_0(r,p)=\eta\,\Theta\left(r_m^2-r^2\right)\Theta\left(p_m^2-p^2\right).
\eeq
%%%%%%%
where $\eta=9/(16\pi^2r_m^3p_m^3)$ is the normalization constant. We will measure all the
lengths in units of $r_m$, which is equivalent to setting $r_m=1$. The VC requires that $2K=-U$, where 
\beq
K=\frac{1}{2}\int \mathrm{d}^3 r\,\mathrm{d}^3 p \,f_0(r,p)\, p^2
\eeq 
is the kinetic energy and 
\beq
U=\frac{1}{2}\int \mathrm{d}^3 r \,\mathrm{d}^3 p \,f_0(r,p) \,\psi_0(r)
\eeq
is the potential energy of the system. The potential $\psi_0(r)$ for the initial waterbag distribution is
%%%%%%%
%\beq\label{eq:psiwb}
%\psi_0(r)=\Theta\left(1-r^2\right)\frac{r^2-3}{2}-\Theta\left(r^2-1\right)\frac{1}{r}.
%\eeq
%%%%%%%
%%%%%%%
\beq\label{eq:psiwb}
\psi_0(r)=
\begin{cases}
\frac{r^2-3}{2} & \text{if } r<1 \\
-\frac{1}{r} & \text{if } r\geq 1.
\end{cases}
\eeq
%%%%%%%
Using Eqs \eqref{eq:wbdist} and \eqref{eq:psiwb} to calculate $K$ and $U$, the VC reduces to $p_m=1$.
In Fig \ref{fig:fel_wb} we plot the joint distribution function $f(\mc{E},\ell^2)$ for
the QSS.
%-------------
%-------------

\begin{figure}
\begin{center}
\includegraphics[width=0.5\textwidth]{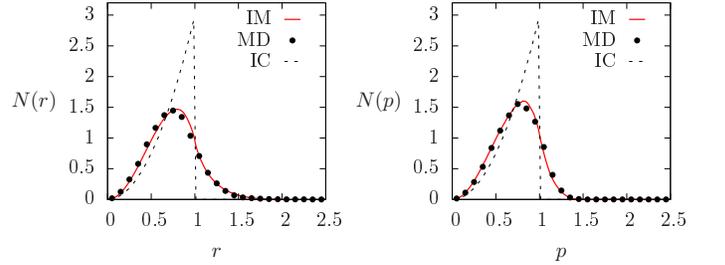}
\caption{Theoretically predicted 
density (left) and momentum (right) distributions (solid lines) for the 
QSS for the initial waterbag distribution.  
The symbols (black dots) are the results of MD simulations. 
The initial $t=0$ density and momentum distributions are plotted with dashed lines -- an initial waterbag distribution is given by Eq \eqref{eq:wbdist}.\label{fig:nrnp}}
\end{center}
\end{figure}

The marginal distribution functions can be calculated using 
Eqs \eqref{eq:nr3} and \eqref{eq:np}  together with Eq \eqref{eq:fel}.
Fig \ref{fig:nrnp} shows the position and velocity distributions $N(r)$ and $N(p)$
predicted by the integrable model.   The symbols are the results of molecular dynamics (MD) simulations. An excellent agreement between the theory and the simulations can be seen.  
%We also compare the energy distribution $f(\mc{E})$ of the IM with the molecular dynamics results, in Fig \ref{fig:fe_wb}. We define the IM energy distribution as
%%%%%%%
%\beq\label{eq:fe_wb}
%f(\mc{E})=\frac{\int\! \dll \,n\!\left(\mc{E},\ell^2\right)}{\int\! \dll\,g\!\left(\mc{E},\ell^2\right)}
%\eeq
%%%%%%%
%where $g\!\left(\mc{E},\ell^2\right)$ and $n\!\left(\mc{E},\ell^2\right)$ are given by Eqs \eqref{eq:ggen} and \eqref{eq:ngen}, respectively.

%-------------
%-------------

\begin{figure}
\begin{center}
\includegraphics[width=0.5\textwidth]{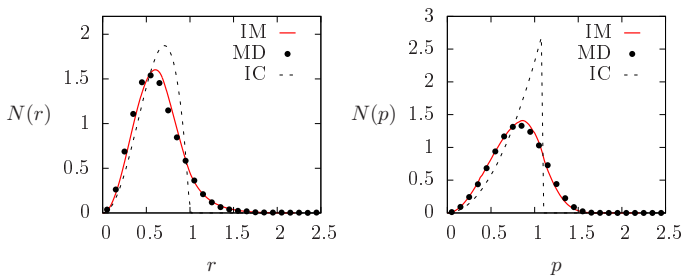}
\caption{Solid lines are the theoretically predicted 
density (left) and momentum (right) distributions for the 
QSS for initial distribution (dashed lines) given by Eq. \eqref{eq:p1dist}.  
The symbols (black dots) are the results of MD simulations. 
\label{fig:nrnp2}}
\end{center}
\end{figure}

One particularly nice feature of the present 
theory is that it can be easily used to predict the final QSS
for any initial distribution as long as it satisfies VC. 
We next study a parabolic initial distribution, given by
%%%%%%%
\beq\label{eq:p1dist}
f_0\left(r,p \right)=\eta\left(1-r^2\right)\Theta\!\left(1-r^2\right)\Theta\!\left(p_m^2-p^2\right)
\eeq
%%%%%%%
with $\eta=45/(32\pi^2 p_m^3)$.
%%%%%%%
%\beq\label{eq:p2dist}
%f(r^2,p^2)=\eta\,\left(\frac{r}{r_m}-\frac{r^2}{r_m^2}\right)\Theta\left(p_m^2-p^2\right)
%\eeq
%%%%%%%
%with $\eta=15/(4\pi^2 p_m^3)$. 
The VC for this distribution is $p_m=5/\sqrt{21}$.  
%and $p_m=\frac{5}{3}\sqrt{17/(42 r_m)}$, respectively.
The marginal distributions predicted by the theory are compared with simulations in Fig \ref{fig:nrnp2}. Once again the agreement is very good.
%the middle row corresponding to an initial distribution given by Eq \eqref{eq:p1dist} and the bottom row to an initial distribution given by Eq \eqref{eq:p2dist}.
%-------------
%-------------
For strongly inhomogeneous
initial distributions, VC is not enough to completely prevent the 
temporal dynamics of the 
mean-field potential.  That is, even if we restrict one moment of 
the distribution function, other moments might still have sufficiently strong dynamics
to excite  parametric resonances.  Indeed, we find that for very strongly
inhomogeneous initial distributions, there is some discrepancy between the theory and
the simulations.  Nevertheless even in these extreme cases the theory remains quite accurate~\cite{SupMatPRL}.

We have presented a theory that is able to predict the particle distribution in the final QSS to
which a
3D self-gravitating system will relax from an initial condition.  
The theory can be used for initial distributions which are isotropic in particle velocity 
and  satisfy the VC.  
%-------------
\begin{figure}
\begin{center}
\includegraphics[width=0.5\textwidth]{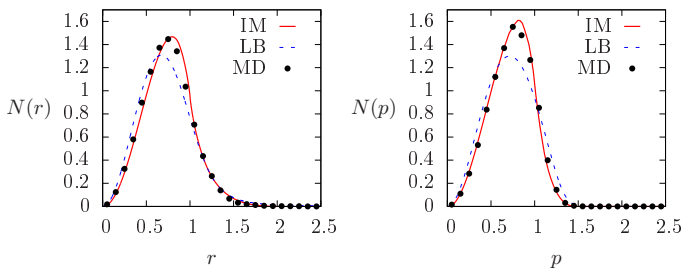}
\caption{ Comparison between the density (left) and momentum (right) distributions calculated using LB statistics and the present theory. Initial distribution is the waterbag in momentum and position, Eq \eqref{eq:wbdist}, satisfying VC. Solid curves are the results of the present theory, 
dashed curves are the predictions of LB theory, and the solid circles 
are the results of MD simulations. \label{figlb}}
\end{center}
\end{figure}
%-------------
It is interesting to compare and contrast our approach with the theory
of violent relaxation developed by Lynden-Bell (LB).  The statistical mechanics of LB is
based on the assumption of ergodicity and perfect mixing of the density levels of the initial 
distribution function over the phase space \cite{Lyn1967}. 
This is contrary to the approach  presented in this Letter which shows
that dynamics of 3D self-gravitating systems with initial distribution satisfying the virial
condition is closer to integrable than ergodic.

Curiously for various systems, in which the particles are either self-bound --- like 1d and 2d gravity --- or are bounded by an external potential or by the topology ---  such as magnetically confined plasmas or spin systems --- the LB approach was found to work
best for the initial waterbag distributions that satisfied the VC~\cite{LevPak2014,BenTel2012}. For
distributions away from VC,  QSS were found to have a characteristic core-halo structure very different from the predictions of LB theory \cite{TelLev2010,TelLev2011,JoyWor2011,PakLev2011,FigRoc2014}.  It was recently observed, however, that for more 
complex inhomogeneous or multilevel
distributions LB theory failed even when the initial distribution function 
satisfied VC \cite{RibBen2014,PakLev2013}.  The failure of 
LB theory can now be attributed to the the almost complete absence of ergodicity and mixing  when the initial distribution
satisfies VC.  The evolution of the
mean-field potential of such systems is almost adiabatic and the dynamics is 
closer to integrable than
to ergodic \cite{RibBen2014}. The relaxation to QSS is the result of phase-mixing of
particles on the same energy shells and not a consequence of ergodicity over the full energy
surface. Indeed for 3D gravitational systems LB theory fails to accurately account for either velocity
or density distributions, as can be seen in Fig. \ref{figlb}, even for the initial virial waterbag distribution,  Eq. \eqref{eq:wbdist}.  Furthermore, LB theory is very difficult to extent to more complicated initial conditions
than a one-level waterbag distribution, while the present approach can, in principle, be used for any arbitrary distribution as long as it satisfies VC.  The goal of the future work will be to extend the
theory presented in this Letter to initial distributions which do not satisfy VC.  
Parametric resonances and particle evaporation, however, make this a very difficult task.

This work was partially supported by the CNPq, FAPERGS, CAPES,
INCT-FCx, and by the US-AFOSR under the grant 
FA9550-12-1-0438.
\bibliography{longrange}

\end{document}